\documentclass{PoS}

\title{In quest of the Yang--Mills vacuum wavefunctional%
\addtocounter{footnote}{1}\thanks{This research was supported in part by the U.S.\ DOE under Grant No.\ DE-FG03-92ER40711 (J.G.), and by the Slovak Grant Agency for Science, Project VEGA No.\ 2/0070/09, by ERDF OP R\&D, Project CE QUTE ITMS~26240120009, and via CE SAS QUTE (\v{S}.O.)}} 

\ShortTitle{In quest of the Yang--Mills vacuum wavefunctional}

\author{Jeff Greensite\\
Physics and Astronomy Dept., San Francisco State University, San Francisco, CA 94132, USA\\
E-mail: \email{jgreensite@gmail.com}}

\author{\addtocounter{footnote}{-2}
				\speaker{\v{S}tefan Olejn{\'\i}k}\\
        Institute of Physics, Slovak Academy of Sciences, SK--845 11 Bratislava, Slovakia\\
        E-mail: \email{stefan.olejnik@savba.sk}}

\abstract{A simple recursion procedure was devised to generate lattice configurations with probability distributions given by simple approximate \YM vacuum wavefunctionals. A few quantities determined in ensembles of these configurations are compared to those computed in configurations generated in standard Monte Carlo simulations of the three-dimensional Yang--Mills theory.}

\FullConference{The many faces of QCD\\
		November 2--5, 2010\\
		Gent, Belgium}

\newcommand\FP{Faddeev--Popov\ }
\newcommand\YM{Yang--Mills\ }
\newcommand\VWF{vacuum wavefunctional\ }

\newcommand\oh{{\textstyle\frac{1}{2}}}

\graphicspath{}

\begin{document}

\section{Introduction}\label{introduction}

	The vacuum of the quantized Yang--Mills gauge field theory contains most information about its distinct features like confinement, chiral symmetry breaking, etc. In the Hamiltonian formulation, this information is carried by the ground-state wavefunctional. In temporal gauge, in $(d+1)$ dimensions, the problem looks very simple: One strives to solve the Schr\"odinger equation
\begin{equation}\label{eq:SchR}
\hat{\cal H}\Psi_0[A]=\displaystyle\int d^d x\left\lbrace-\oh
\frac{\delta^2}{\delta A_k^a(x)^2}+{\textstyle\frac{1}{4}}F_{ij}^a(x)^2\right\rbrace\Psi_0[A]=E_0\Psi_0[A]
\end{equation}
with an additional constraint that physical states are invariant under infinitesimal local gauge transformations (Gau\ss' law):
\begin{equation}\label{eq:GaussLaw}
\displaystyle\left(\delta^{ac}\partial_k+
g\epsilon^{abc}A_k^b\right)
\frac{\delta}{\delta A_k^c}\Psi[A]=0.
\end{equation}
Subtleties of this problem were outlined more than 30 years ago by one of us (J.G.) \cite{Greensite:1979yn}. It was argued that at large distance scales one expects the wavefunctional to assume the so called \emph{dimensional-reduction form}:
\begin{equation}\label{eq:dim_reduction}
\Psi_0^{\mathrm{eff}}[A] \approx\exp\left[-\mu\int d^dx\; F^a_{ij}(x)F^a_{ij}(x)\right],
\end{equation}
i.e.\ a vacuum with color-magnetic fields fluctuating independently in each spacetime point. In such a case the computation of a spacelike loop in $(d+1)$ dimensions reduces to the calculation of a Wilson loop in Yang--Mills theory in $d$ Euclidean dimensions. If the property existed for \YM theories $(3+1)$ and $(2+1)$ dimensions, then these would be confining, since the theory in 2 euclidean dimensions exhibits the area law.\footnote{In the rest of this paper, we will discuss exclusively the case of $(2+1)$ dimensions.} However, the dimensional-reduction form cannot be the whole story, it does not provide correct short-distance structure of the theory. 

	As a step forward, Greensite \cite{Greensite:1979ha} proposed a systematic strong-coupling expansion of the \YM \VWF in the form:
\begin{equation}\label{eq:exp_R}
\Psi_0[U]={\cal{N}}\exp(R[U]),
\end{equation}
where the function $R$ in the exponential is an expansion in terms of closed loops -- products of link variables $U$ along closed contours on the lattice:
\begin{equation}\label{eq:strong_coupling}
\centerline{\includegraphics[width=0.4\textwidth]{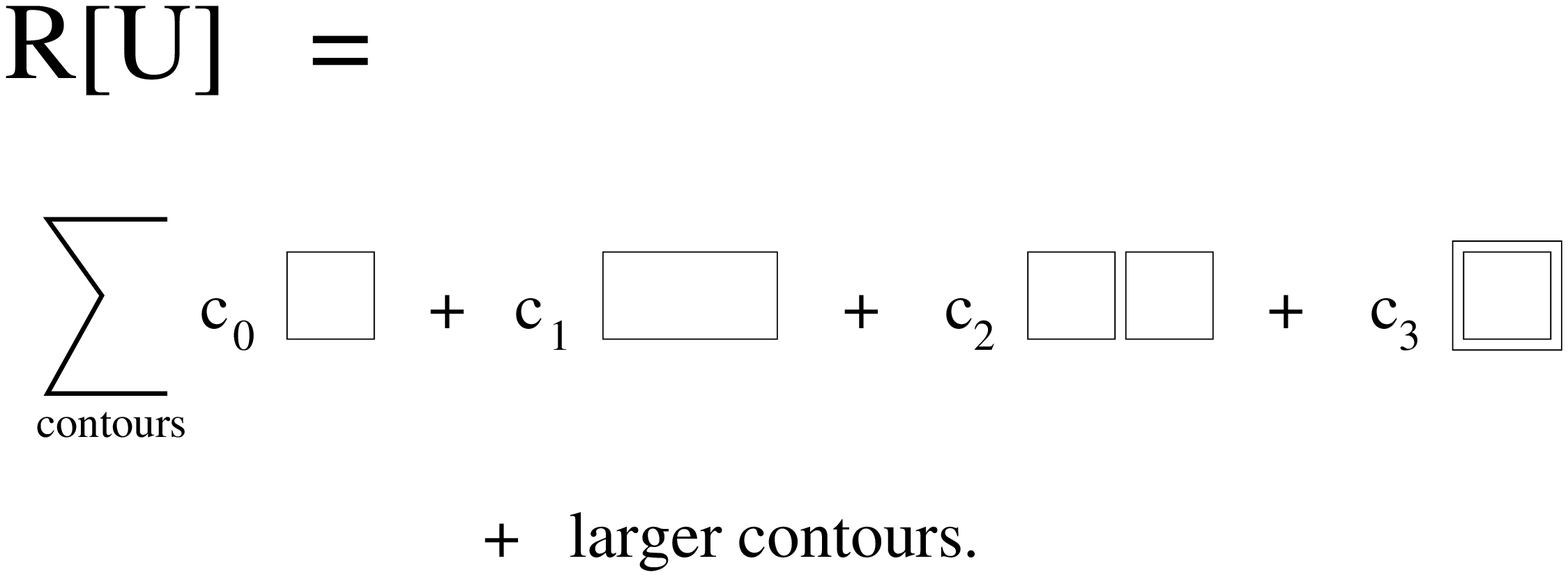}}
\end{equation}
It was later shown by Guo, Chen and Li \cite{Guo:1994vq} that for smoothly varying gauge fields the first terms of the expansion are expressed through the color magnetic field strength, $B^a(x)=F_{12}^a(x)$,
and the covariant laplacian in the adjoint representation, ${\cal D}^2={\cal D}_k\cdot{\cal D}_k$, where ${\cal D}_k[A]$ denotes the covariant derivative in the adjoint representation:
\begin{equation}\label{eq:Guo_expansion}
R[U]\propto
-\frac{1}{\beta}\left( a\kappa_0\mbox{Tr}\;[B^2]-a^3
\kappa_2\mbox{Tr}\;[B(-{\cal D}^2)B]+\dots\right),
\end{equation}
where
\begin{equation}\label{Guo_constants}
\kappa_0=\oh c_0+2(c_1+c_2+c_3),\quad \kappa_2={\textstyle\frac{1}{4}}c_1\qquad\mbox{with}\qquad
c_0={\cal O}(\beta^2),\quad c_1, c_2, c_3={\cal O}(\beta^4).
\end{equation}

	With only a few exceptions (see references in \cite{Greensite:in-prep}), there was not much work done in this area after the initial efforts. However, recently interest in the problem of the \YM \VWF has been revived and various plausible proposals for the vacuum states have been advanced. We will present tests of two of them which differ by their motivation but share some common features. More results and other proposals will be covered elsewhere \cite{Greensite:in-prep}. 

\section{Two proposals}\label{proposals}

	With a grain of imagination one can assume that an expansion of the form (\ref{eq:Guo_expansion}) might result from a \VWF 
\begin{equation}\label{eq:generalVWF}
\Psi_0[A]{=}{\cal{N}}\exp\left[-\oh\int d^2xd^2y\; 
B^a(x)\;{K^{ab}_{xy}[-{\cal{D}}^2]}\;B^b(y)\right]
\end{equation}
with a kernel depending on the gauge coupling $g$ and the adjoint covariant laplacian. That kernel cannot be arbitrary -- one should be able to reproduce the QED vacuum wavefunctional [for SU(2) in $(2+1)$-dimensional spacetime] for $g\to 0$, which is known to be \cite{Wheeler}:
\begin{equation}\label{eq:QED_VWF}
\Psi_0[A]\;
{\propto}\;
\exp\left\{-\oh\displaystyle\int d^2xd^2y\; 
[\nabla\times A(x)]{\displaystyle
\left(\frac{1}{\sqrt{-\nabla^2}}\right)_{xy}}[\nabla\times A(y)]\right\rbrace.
\end{equation}
The kernel has thus to satisfy the condition: 
\begin{equation}\label{QED_condition}
\lim_{g\to 0} K^{ab}_{xy}[-{\cal{D}}^2]=\left(\frac{\delta^{ab}}{\sqrt{-\nabla^2}}\right)_{xy}.
\end{equation}

	The above condition is, by construction, satisfied by the approximate \VWF proposed recently by the present authors \cite{Greensite:2007ij}:
\begin{equation}\label{GO}
\Psi_{\mathrm{GO}}[A]{=}{\cal{N}}
\exp\left[-\oh\displaystyle\int d^2xd^2y\; 
B^a(x){\displaystyle
\left(\frac{1}{\sqrt{(-{\cal D}^2-
\lambda_0)+m^2}}\right)_{xy}^{ab}}B^b(y)\right].
\end{equation}
Here $\lambda_0$ denotes the lowest eigenvalue of $(-{\cal D}^2)$, and $m$ is a constant (mass) parameter proportional to $g^2\sim 1/\beta$. It is similar to a proposal by Samuel \cite{Samuel:1996bt}; the difference lies in the subtraction of $\lambda_0$ which is crucial, since $(-{\cal D}^2)$ has a positive-definite spectrum and we have hints that its lowest eigenvalue diverges in the continuum limit.

	The supporting evidence for this proposal comes from four sources \cite{Greensite:2007ij,Greensite:2010tm}:
%

	1.\ The proposed form is a good approximation to the true vacuum for strong fields constant in space and varying only in time.

	2.\ If we divide the magnetic field strength $B(x)$ into ``fast'' and ``slow'' components, the part of the vacuum wavefunctional that depends on $B_\mathrm{slow}$ takes on the dimensional-reduction form. The fundamental string tension is then easily computed as 
\begin{equation}\label{eq:sigma_F}
\sigma_\mathrm{f}= 3mg^2/16.
\end{equation}
%

	3.\ If one takes the mass $m$ in the wavefunctional as a free variational parameter and computes (approximately) the expectation value of the \YM hamiltonian, one finds that a non-zero (finite) value of $m$ is energetically preferred.

	4.\ Results for the mass gap, and the Coulomb-gauge ghost propagator and the color-Coulomb potential computed from $\Psi_{\mathrm{GO}}[A]$ are in good agreement with those derived from standard Monte Carlo simulations (see below).

	Despite this evidence, our proposal represents only an educated guess, motivated by the form of the QED vacuum state, dimensional reduction, and gauge invariance. 
	
	A more sophisticated approach has been followed in $(2+1)$ dimensions by Karabali, Nair and collaborators \cite{Karabali:1998yq}. In the temporal gauge ($A_0=0$) they combine the remaining two components of the gauge potential into complex fields:
\begin{equation}\label{eq:complex}
\mathbf{A}\equiv\oh\left(\mathbf{A}_1+i \mathbf{A}_2\right),
\qquad
\bar\mathbf{A}\equiv\oh\left(\mathbf{A}_1-i \mathbf{A}_2\right),
\end{equation}
and then introduce new variables, a matrix-valued field $\mathbf{M}\in \mathrm{SL}(N,\mathcal{C})$, which is related to $\mathbf{A}, \bar\mathbf{A}$ via
\begin{equation}\label{eq:KKN_parametrization}
\mathbf{A}=-(\partial_z\mathbf{M})\mathbf{M}^{-1},
\qquad
\bar\mathbf{A}={\mathbf{M}^\dagger}^{-1}(\partial_{\bar{z}}\mathbf{M}^\dagger),
\end{equation}
where $z=x_1-i x_2$ and $\bar{z}=x_1+i x_2$ are the usual holomorphic variables in the complex plane.

	Under a gauge transformation $\Omega$, $\mathbf{M}$ transforms covariantly, $\mathbf{M}\to\Omega\mathbf{M}$, and can be used to define gauge-invariant field variables:
\begin{equation}\label{eq:KKN_variables}
\mathbf{H}\equiv \mathbf{M}^\dagger\mathbf{M}, \qquad J^a\sim\mbox{Tr}\left(T^a(\partial\mathbf{H})\mathbf{H}^{-1}\right),
\end{equation}
through which one can express the hamiltonian, inner products of physical states, and the vacuum wavefunctional.

	Karabali et al.\ argue that the part bilinear in field variables $J^a$, when expressed in usual variables, has the form:
\begin{equation}\label{KKN}
\Psi_{\mathrm{KKN}}[A] \approx
\exp\left[-\oh\int d^2xd^2y\; 
B^a(x){\displaystyle
\left(\frac{1}{\sqrt{-{\nabla}^2+m^2}+m}\right)_{xy}^{ab}}B^b(y)\right],
\end{equation}
which, however, is not gauge-invariant. One can imagine that higher-order terms in $J^a$ might convert the ordinary laplacian in Eq.\ (\ref{KKN}) into the covariant laplacian, leading to:
\begin{equation}\label{KKN2}
\Psi_{\mathrm{KKN}'}[A] \approx \exp\left[-\oh\int d^2xd^2y\; 
B^a(x){\displaystyle
\left(\frac{1}{\sqrt{-{\cal D}^2+m^2}+m}\right)_{xy}^{ab}}B^b(y)\right].
\end{equation}
This form is still hardly sustainable, because of the divergence of the lowest eigenvalue $\lambda_0$ of the adjoint covariant laplacian (discussed above).

	Instead of the form (\ref{KKN2}) we will subject to lattice tests a ``KKN-inspired'' or ``hybrid'' wavefunctional which has (\ref{KKN}) and (\ref{KKN2}) as starting point and agrees with them for abelian gauge configurations, but in which, similarly to the GO proposal, the covariant laplacian $(-{\cal D}^2)$ is replaced by the subtracted one $(-{\cal D}^2-\lambda_0)$:   
\begin{equation}\label{hybrid}
\Psi_{\mathrm{hybrid}}[A]{=}{\cal{N}}
\exp\left[-\oh\displaystyle\int d^2xd^2y\; 
B^a(x){\displaystyle
\left(\frac{1}{\sqrt{(-{\cal D}^2-
\lambda_0)+m^2}+m}\right)_{xy}^{ab}}B^b(y)\right].
\end{equation}

\section{Tests of the proposals}\label{tests}

	Our aim is to test how good/bad is the approximation of the true \YM \VWF by the proposed approximate forms, Eq.\ (\ref{GO}) and (\ref{hybrid}). To achieve this goal, we take a set of operators $\{\hat{Q}[A]\}$ that depend on gauge fields $A$, and compute (and compare) their expectation values:

$\bullet$\ {$\langle\Psi_0^\mathrm{true}\vert\hat{Q}[A]\vert\Psi_0^\mathrm{true}\rangle=\langle\;Q[A]\;\rangle_\mathrm{{MC}}$} in \textbf{\textit{Monte Carlo lattices}}, i.e.\ an ensemble of two-dimensional slices of configurations generated by MC simulations of the three-dimensional euclidean SU(2) lattice gauge theory with standard Wilson action at a coupling $\beta_\mathrm{E}$; from each configuration, only one (random) slice at fixed euclidean time is taken; 

$\bullet$\ {$\langle\Psi_0\vert\hat{Q}[A]\vert\Psi_0\rangle=\langle\;Q[A]\;\rangle_\mathrm{{recursion}}$} in \textbf{\textit{``recursion'' lattices}}, i.e.\ an ensemble of independent two-dimensional lattice configurations generated with the probability distribution given by a proposed vacuum wavefunctional, with parameters $m$ and $g^2$ fixed to some reasonable values, to be able to compare to the MC ensemble.

\paragraph{\underline{Numerical simulation of $\vert\Psi_0\vert^2$}}\label{simulation}

	The generation of recursion lattices whose probability distribution $P[A]$ is given by the square of a wavefunctional of the type (\ref{eq:generalVWF}) is based on the following idea~\cite{Greensite:2007ij}: Define a probability distribution for gauge fields $A$ with the kernel $K$ controlled by an independent ``background'' configuration~$A'$
\begin{equation}\label{eq:background_P}
{\cal P}[A;K[A']]={\cal N}\exp\left[-\displaystyle\int d^2xd^2y\; 
B^a(x;A){K_{xy}^{ab}[A']}B^b(y;A)\right],
\end{equation}
where the field strength $B$ is computed from $A$, and both $A$ and $A'$ are fixed to an appropriate gauge. If the variance of the kernel $K[A]$ in the probability distribution $P[A]$ is small after the choice of gauge, then one can write down a chain of approximate relations:
\begin{equation}\label{approximate_P}
{P[A]}={\cal P}[A;K[A]]\approx{\cal P}[A;\langle K\rangle]\displaystyle
={\cal P}\left[A;\int dA'\;{\cal P}[A;K[A']] {P[A']}\right]
\approx\int dA'\;{\cal P}[A;K[A']] {P[A']}.
\end{equation}
The probability distribution $P[A]$ can then obtained by solving (\ref{approximate_P}) iteratively: 
\begin{equation}
{P^{(1)}[A]={\cal P}[A;K[0]]},\quad\dots,\quad
{P^{(k+1)}[A]=\displaystyle\int
dA'\;{\cal P}[A;K[A']] P^{(k)}[A']}.
\end{equation}

	Practical implementation of the recursion procedure consists of the following steps: Choose $A_1=0$ (axial gauge) and $A_2\ne 0$, then
	
	(i) given $A_2$, set $A'_2=A_2$,
	
	(ii) ${\cal P}\left[A;{{K}}[A']\right]$ is gaussian in $B$, diagonalize ${{K}}[A']$ and generate a new $B$-field stochastically,
	
	(iii) from $B$ calculate $A_2$ in axial gauge and compute everything of interest,
	
	(iv) go back to step (i), repeat as many times as necessary.
	
	The procedure converges rapidly, one needs ${\cal O}(10)$ cycles above, and the assumption about a small variance of $K$ among configurations is supported a posteriori by the
absence of large fluctuations of the spectrum of $K$ evaluated on individual recursion lattices.

\paragraph{\underline{Choice of \VWF parameters $m$ and $g$}}\label{choice}

	Our next task is to select appropriate values for parameters $g$ (or $\beta=4/g^2$) and $m$ of recursion lattices to be able to compare to MC lattices with Wilson-action coupling $\beta_\mathrm{E}$. In our earlier studies \cite{Greensite:2007ij,Greensite:2010tm} with $\Psi_\mathrm{GO}$ we chose $\beta=\beta_\mathrm{E}$, and fixed $m$ using Eq.\ (\ref{eq:sigma_F}) by the measured value of the fundamental string tension (in lattice units): $m(\beta_\mathrm{E},L)=4\beta_\mathrm{E}\sigma_\mathrm{f}(\beta_\mathrm{E},L)/3$.
	
	Another possibility is to use for fixing $\beta$ and $m$ some information about the true \YM \VWF at $\beta_\mathrm{E}$ for a set of simple gauge-field configurations. The square of the vacuum wavefunctional for some trial configurations (non-abelian constant fields, abelian or non-abelian plane waves) can be computed numerically in simulations of the three-dimensional \YM theory. Take a set of time-independent configurations ${\cal U}=\{U^{(j)}(\mathbf{x}), j=1, \dots, M\}$. The method (proposed long ago~\cite{Greensite:1988rr} and described in more detail in \cite{Greensite:in-prep}) is based on the following identity:
\begin{equation}\label{eq:Psi_2}
\vert\Psi[U^{(j)}]\vert^2=\displaystyle\frac{1}{Z}\int [DU]\delta(U_0)\prod_\mathbf{x} \delta\left[U(\mathbf{x},0)-U^{(j)}(\mathbf{x})\right]e^{-S}.
\end{equation}
In practice, one measures the probability in a modified lattice Monte Carlo simulation: The links at $t=0$ are constrained to belong to a configuration from the set ${\cal U}$. In a MC update all links, except those with $t=0$, are updated by the usual heat bath method. On the $t=0$ slice, one of the $M$ configurations from the set ${\cal U}$ is selected at random, and then accepted/rejected by the Metropolis algorithm. Len $N_j$ denote the total number of times that the $j$-th configuration from the set is accepted, and $N_\mathrm{tot}$ the total number of updates of the $t=0$ plane. Then:
\begin{equation}\label{eq:measured_Psi_2}
\vert\Psi[U^{(j)}]\vert^2\propto\lim_{N_\mathrm{tot}\ \mathrm{large}}\frac{N_j}{N_\mathrm{tot}}.
\end{equation}

	For determining \VWF parameters we measured probabilities of abelian plane waves with fixed (maximal) wavelength $\lambda=L$ and varying amplitudes:
\begin{eqnarray}
&&U_1^{(j)}(n_1,n_2)=\sqrt{1-a_j(n_2)^2}\mathbf{1}_2+i a_j(n_2)\sigma_3,
\qquad
U_2^{(j)}(n_1,n_2)=\mathbf{1}_2,\\
\mbox{with }&&a_j(n_2)=\sqrt{\frac{\alpha+\gamma j}{L^2}}\cos\frac{2\pi n_2}{L},
\qquad
p^2=2\left(1-\cos\frac{2\pi}{L}\right).
\end{eqnarray}
The probabilities measured in the Monte Carlo simulation described above can be parametrized by
\begin{equation}
\vert\Psi_\mathrm{MC}[U^{(j)}]\vert^2=\exp(-R_\mathrm{MC}[U^{(j)}]-R_0),\qquad
R_\mathrm{MC}[U^{(j)}]=2(\alpha+\gamma j)\;\omega_\mathrm{MC}(p)+\mbox{ const}.
\end{equation}
Similarly, for a theoretical Ansatz of vacuum wavefunctional:
\begin{equation}\label{eq:R_ansatz}
R_\mathrm{Ansatz}[U^{(j)}]=2(\alpha+\gamma j)\;\omega_\mathrm{Ansatz}(p)+\mbox{ const.}
\end{equation}
For the proposals discussed in Section \ref{proposals}:\footnote{For abelian configurations, the KKN (\ref{KKN}) and hybrid (\ref{hybrid}) wavefunctionals coincide.} 
\begin{equation}\label{omegas}
\omega_\mathrm{{GO}}(p)=\frac{1}{g^2}\frac{p^2}{\sqrt{p^2+m^2}},
\qquad
\omega_\mathrm{{KKN}}(p)=\frac{1}{g^2}\frac{p^2}{\sqrt{p^2+m^2}+m}.
\end{equation}

\begin{figure}[t!]	
\centerline{\includegraphics[width=0.45\textwidth]{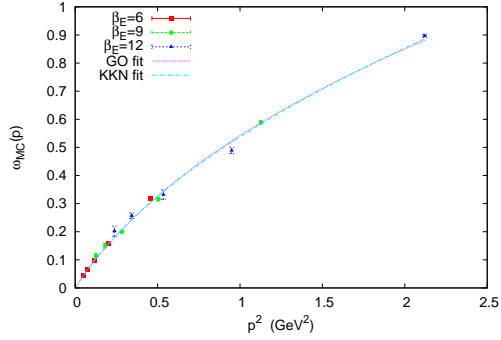}}
\caption{Cumulative data for $\omega_{\mathrm{MC}}$ vs.\ $p^2$ in physical units, on lattices of extensions
$L=16,24,32,40,48$, and euclidean lattice couplings $\beta_\mathrm{E}=6,9,12$.  The curves (hardly distinguishable from one another) represent $\omega_{\mathrm{GO}}(p)$ and $\omega_{\mathrm{KKN}}(p)$ for fitted parameters $m_\mathrm{phys}$ and $g^2_\mathrm{phys}$ given in Table~\protect\ref{tab:physical_parameters}.}\label{fig:all-omega}
\end{figure}

	Fig.\ \ref{fig:all-omega} shows results for $\omega_\mathrm{MC}$ vs.\ $p^2$ in physical units for a number of $\beta_\mathrm{E}$ values and lattice sizes~$L$. The scale was set by the conventional value of $(0.44\mbox{ GeV})^2$ for the physical string tension, i.e.\ the lattice spacing is $a(\beta_\mathrm{E},L)={\sqrt{{\sigma_\mathrm{f}(\beta_\mathrm{E},L)}}}/(0.44\ \mathrm{GeV})$. The data were fitted by functional forms in Eq.\ (\ref{omegas}), the resulting parameters are summarized in Table~\ref{tab:physical_parameters}.

\begin{floatingtable}[r]
\centering
\begin{tabular}{| c  | c c |}
\hline 
variant & {$m_\mathrm{phys}$} & {$g^2_\mathrm{phys}$} \\
\hline
{GO}   & 0.771 & 1.465\\
{KKN} & 0.420 & 1.237\\\hline
\end{tabular}
\caption{Vacuum wavefunctional parameters (phys.\ units).}\label{tab:physical_parameters}
\end{floatingtable}

	Parameters of the proposed wavefunctionals in lattice units were fixed using:
\begin{equation}
{g^2}(\beta_{\mathrm{E}},L)=g^2_\mathrm{phys}\;{a(\beta_\mathrm{E},L)},\qquad
m(\beta_{\mathrm{E}},L)=m_\mathrm{phys}\;{a(\beta_\mathrm{E},L)}.
\end{equation}
We will present below results for $\beta_\mathrm{E}=9$, the actual parameter values used at this coupling are listed in Table~\ref{tab:lattice_parameters}.\footnote{The results obtained for the GO \VWF with two variants of fixing its parameters, described in the text, almost do not differ \cite{Greensite:2010yp}.}

\begin{table}[b!]
\centering
\begin{tabular}{| c c | c c | c c | c c |}
\hline
& & & & \multicolumn{2}{c |}{GO} & \multicolumn{2}{c |}{hybrid}\\
\cline{5-6}\cline{7-8}
${\beta_\mathrm{E}}$ & ${L}$ & $\sqrt{{\sigma_\mathrm{f}(\beta_\mathrm{E},L)}}$ & ${a(\beta_\mathrm{E},L)}$ & ${\beta}$  & ${m}$ & ${\beta}$ & ${m}$ \\\hline
9 & 32 & 0.162 & 0.367 & 7.43 & 0.283 & 8.80 & 0.154\\\hline
\end{tabular}
\caption{Values of $\beta$ and $m$ for the GO and hybrid wavefunctionals at $\beta_\mathrm{E}=9$, derived from the parameters in Table \protect\ref{tab:physical_parameters}. (The value of $\sqrt{{\sigma_\mathrm{f}(\beta_\mathrm{E},L)}}$ comes from Monte Carlo simulations of the standard Wilson action in three Euclidean dimensions \protect\cite{Meyer:2003wx}.)}
\label{tab:lattice_parameters}
\end{table}

\paragraph{\underline{Results}}\label{results}

	We focus on two important quantities defined in Coulomb gauge, the Coulomb-gauge ghost propagator:
\begin{equation}\label{ghost_propagator}
G(R)=\left.\left\langle\left({\cal{M}}[A]^{-1}\right)^{aa}_{xy}\right\rangle\right\vert_{\vert x-y\vert=R}=
\displaystyle\left.\left\langle\left(-\frac{1}{\nabla\cdot{\cal{D}}[A]}\right)^{aa}_{xy}\right\rangle\right\vert_{\vert x-y\vert=R}
\end{equation}
and the color-Coulomb potential:
\begin{equation}\label{Coulomb_potential}
V(R)\propto\left.-\left\langle\left({\cal{M}}[A]^{-1}(-\nabla^2){\cal{M}}[A]^{-1}\right)^{aa}_{xy}\right\rangle\right\vert_{\vert x-y\vert=R}
=\displaystyle\left.-\left\langle\left(\frac{1}{\nabla\cdot{\cal{D}}[A]}(-\nabla^2)\frac{1}{\nabla\cdot{\cal{D}}[A]}\right)^{aa}_{xy}\right\rangle\right\vert_{\vert x-y\vert=R}.
\end{equation}
It was argued by Gribov \cite{Gribov:1977wm} and Zwanziger \cite{Zwanziger:1998ez}, that the	low-lying spectrum of the \FP operator,  ${\cal{M}}[A]=-\nabla\cdot{\cal{D}}[A]$, in Coulomb gauge probes properties of nonabelian gauge fields that are crucial for the confinement mechanism. The ghost propagator in Coulomb gauge and the color-Coulomb potential are directly related to the inverse of the \FP operator, and play a role in various confinement scenarios. In particular, the color-Coulomb potential represents an upper bound on the physical potential between a static quark and antiquark \cite{Zwanziger:2002sh}.

	An important point to mention is the equality of the vacuum wavefunctionals in temporal and Coulomb gauge (see e.g.~\cite{Greensite:2004ke}), when evaluated on gauge fields satisfying the Coulomb gauge condition $\nabla\cdot A=0$, and which lie in the first Gribov region. Our numerical method described above generates configurations in the temporal gauge, these are then transformed to Coulomb gauge, and Coulomb-gauge observables are evaluated in the transformed configurations.

\begin{figure}[t!]
\begin{minipage}[t]{0.49\linewidth}	
\centerline{\includegraphics[width=\textwidth]{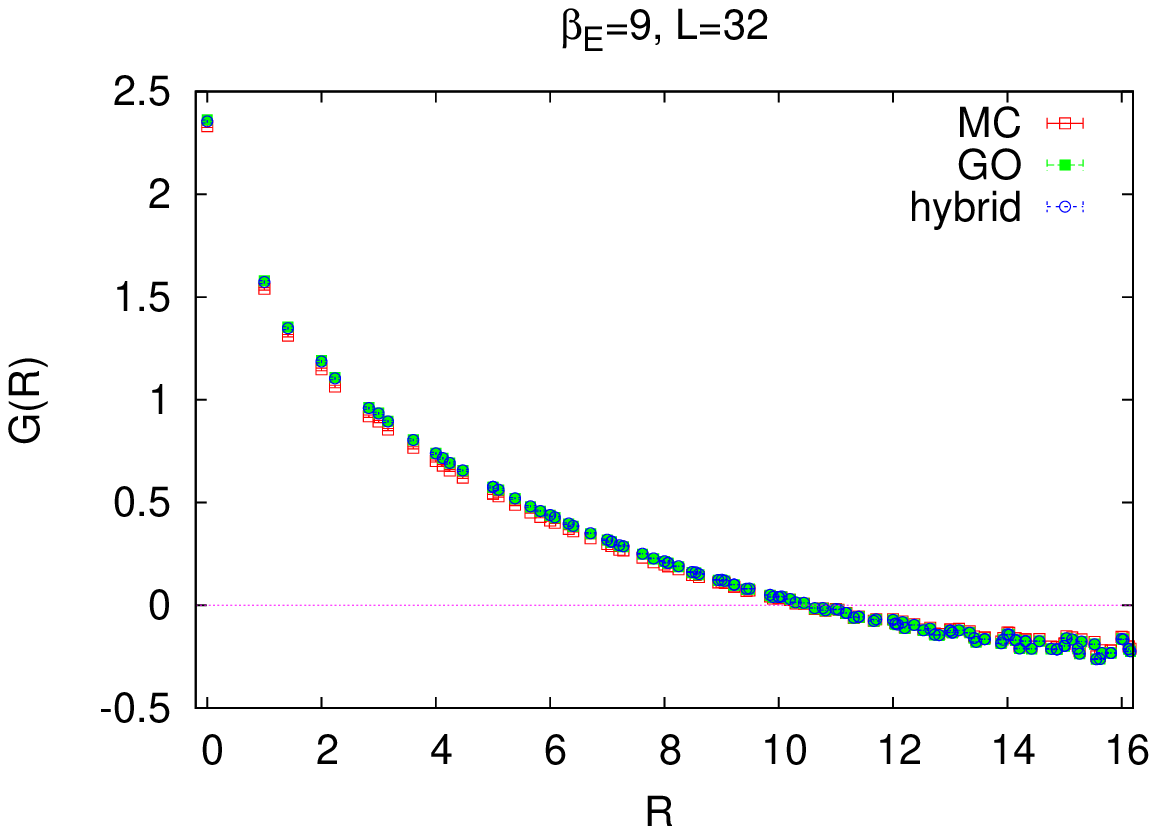}}
\caption{The Coulomb-gauge ghost propagator.}\label{fig:propagator_b9_l32_all}
\end{minipage}\hspace{0.02\linewidth}
\begin{minipage}[t]{0.49\linewidth}
\centerline{\includegraphics[width=\textwidth]{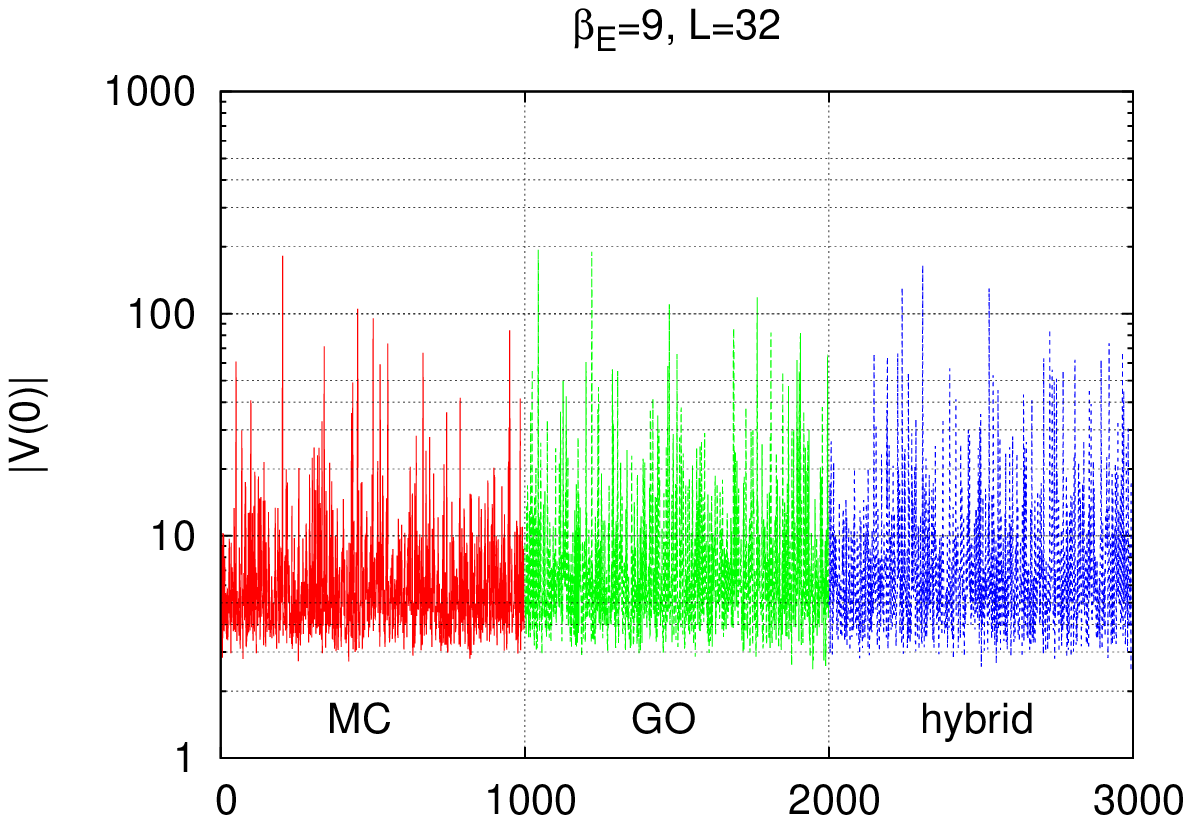}}
\caption{$\vert V(0)\vert$ in the individual configurations.}\label{fig:v0_b9_l32_with_KKN}
\end{minipage}
\end{figure}

	Figure \ref{fig:propagator_b9_l32_all} displays the equal-time ghost propagator in Coulomb gauge computed in a standard Monte Carlo simulation on a $32^3$ lattice at $\beta_\mathrm{E}=9$, together with results obtained from recursion lattices with probability distributions given by $\Psi^2_\mathrm{GO}$ and $\Psi^2_\mathrm{hybrid}$, generated using $\beta$ and $m$ values listed in Table~\ref{tab:lattice_parameters}. The agreement is quite perfect, for all three ensembles.
	
\begin{figure}[t!]	
\begin{tabular}{c c}
\includegraphics[width=0.45\textwidth]{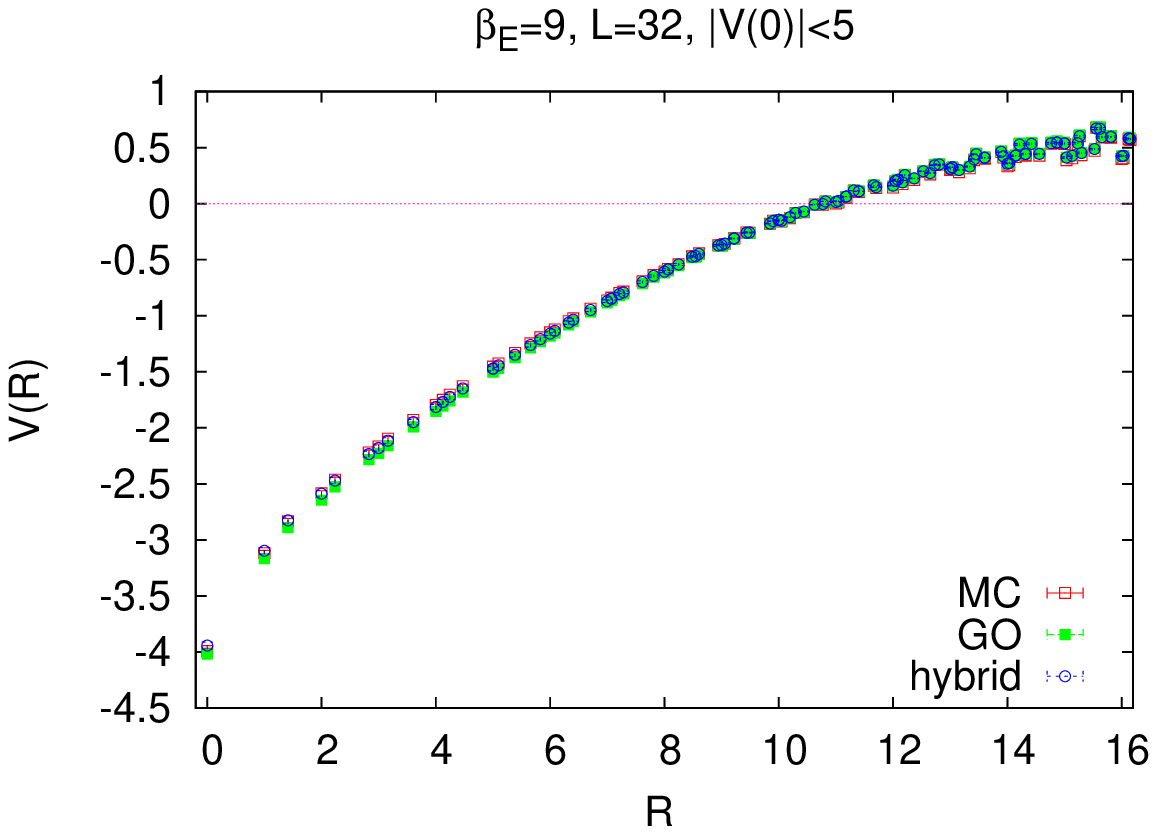}&
\includegraphics[width=0.45\textwidth]{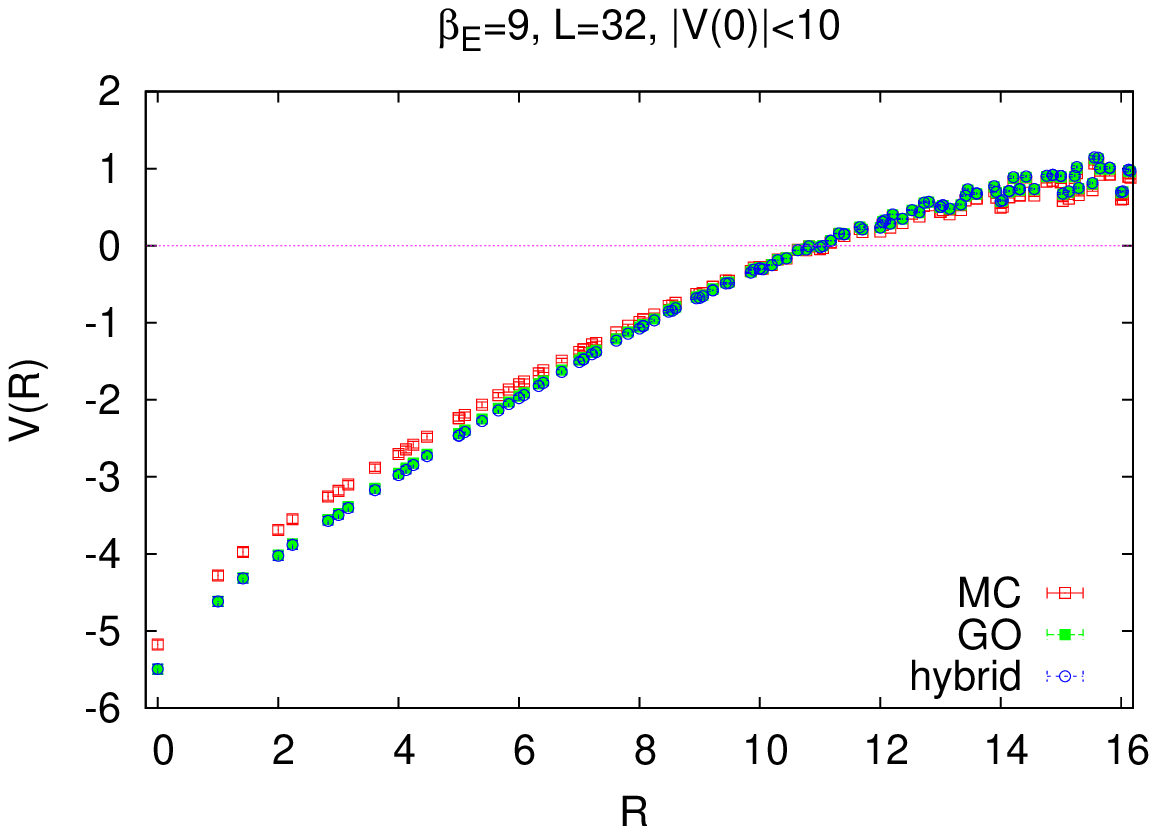}\\
\includegraphics[width=0.45\textwidth]{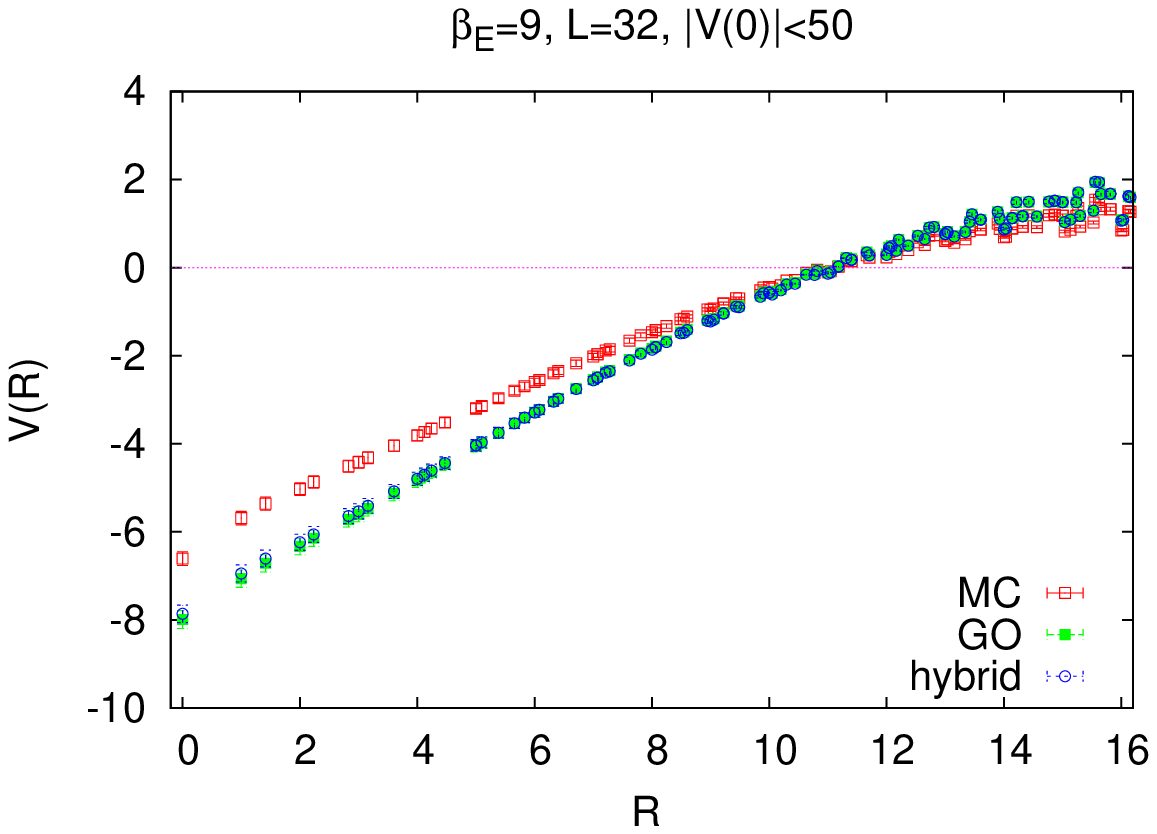}&
\includegraphics[width=0.45\textwidth]{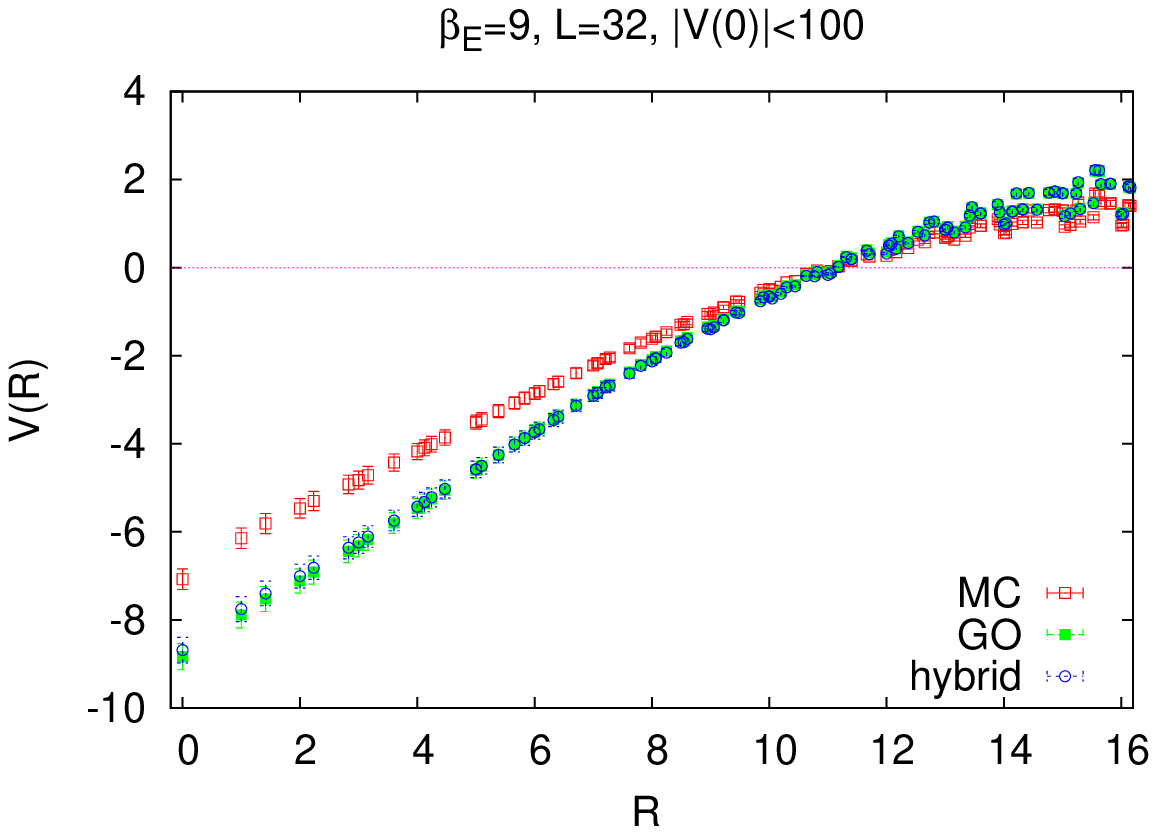}
\end{tabular}
\caption{The color-Coulomb potential for illustrative values of the cut $\kappa$.}\label{fig:v0_b9_l32_with_limits}
\end{figure}

	Figure \ref{fig:v0_b9_l32_with_KKN} is crucial for understanding results for the color-Coulomb potential, which are shown in Fig.~\ref{fig:v0_b9_l32_with_limits}. One can evaluate the potential in each individual lattice configuration. Figure~\ref{fig:v0_b9_l32_with_KKN} displays values of $\vert V(0)\vert$ in MC lattices, GO and hybrid recursion lattices. It is clearly seen that most configurations have $\vert V(0)\vert$ in the range between about 2 and 10 (about 80\%), but there are rare instances of configurations with much higher values. These ``exceptional'' lattices possess a still positive, but very small value of the lowest eigenvalue of the \FP operator, and are therefore rather difficult do gauge-fix to Coulomb gauge. If we discard from our ensembles configurations with $\vert V(0)\vert$ greater than some cut $\kappa$, we obtain results illustrated in Fig.~\ref{fig:v0_b9_l32_with_limits}. For $\kappa=5$ or even 10 the agreement of potentials for MC, GO, and hybrid lattices is reasonable. However, as the cut is increased, the agreement deteriorates. The GO and hybrid potentials are still roughly linear (and hardly distinguishable from each other), but deviate quantitatively from the MC result. This indicates a discrepancy in the tails of the probability distributions. While the ghost propagator (\ref{ghost_propagator}), containing only a single factor of the inverse \FP operator, is rather insensitive to the tails and its values are mainly determined by the bulk of configurations, the color-Coulomb potential (\ref{Coulomb_potential}) involves two factors and is more sensitive to the tails.
	
	But what makes the probability distributions corresponding to $\Psi^2_\mathrm{GO}$ and $\Psi^2_\mathrm{hybrid}$ so close? We believe that the reason is that both wavefunctionals have -- for optimal choices of their parameters -- about the same dimensional-reduction limit, and the results are mainly sensitive to that limit, not to the detailed functional form of the kernel $K$ that enters the wavefunctional. 
	
\section{Summary}\label{conclusions}

1.\ We described a recursion procedure that allows to generate lattice configurations with probability distributions given by approximate \YM vacuum wavefunctionals of certain simple forms like (\ref{GO}) and (\ref{hybrid}).

2.\ Relative magnitudes of the true vacuum wavefunctional on particular sets of configurations (abelian plane waves, non-abelian constant configurations) can be computed numerically.

3.\ Parameters of approximate vacuum wavefunctionals can be fixed e.g.\ by fitting the results for long-wavelength abelian plane waves.

4.\ The two tested proposals (GO, KKN-inspired hybrid) provide Coulomb-gauge quantities almost indistinguishable, and in reasonable agreement with lattice Monte Carlo results (with some discrepancy in color-Coulomb potentials).

5.\ GO and hybrid vacuum wavefunctionals seem to agree with the true Yang--Mills vacuum wavefunctional for the bulk of the probability distribution.

6.\ The important common property of both tested approximate vacuum wavefunctionals appears to be their almost identical dimensional-reduction form.

Only a subset of our recent results was covered in the present contribution. The interested reader should consult Refs.\ \cite{Greensite:in-prep,Greensite:2007ij,Greensite:2010tm,Greensite:2010yp} for additional details and more data.

\medskip
\noindent
{\v S}.O.\ is grateful to the organizers, in particular David Dudal and Nele Vandersickel, for creating very stimulating informal atmosphere of the workshop in the beautiful setting of the city of Ghent.

\end{document}